\documentclass[aps,twocolumn,showpacs,amsmath,amssymb,prb]{revtex4}
\usepackage{graphicx}
\usepackage{amssymb,amsmath,latexsym}
\usepackage{bm}
\usepackage{ulem}
\usepackage{bm,color}

\begin{document}

\title{Structural instability and the Mott-Peierls transition
in a half-metallic hollandite : K$_{2}$Cr$_{8}$O$_{16}$}

\author{Sooran Kim}
\author{Kyoo Kim}
\email[ ]{kyoo@postech.ac.kr}
\author{B. I. Min}
\email[ ]{bimin@postech.ac.kr}
\affiliation{Department of physics, PCTP,
Pohang University of Science and Technology, Pohang, 790-784, Korea}

\begin{abstract}
In order to explore the driving mechanism of the concomitant metal-insulator
and structural transitions in 
quasi-one-dimensional
hollandite K$_{2}$Cr$_{8}$O$_{16}$,
electronic structures and phonon properties are investigated
by employing the {\it ab initio} density functional theory (DFT) calculations.
We have found that the 
imaginary phonon frequency reflecting the structural instability
appears only in the DFT+$U$ ($U$: Coulomb correlation) calculation,
which indicates that the Coulomb correlation plays an essential
role in the structural transition.
The lattice displacements of the softened phonon at X explain
the observed lattice distortions in K$_{2}$Cr$_{8}$O$_{16}$ perfectly well,
suggesting 
the Peierls distortion vector {\bf Q} of X (0, 0, 1/2).
The combined study of electronic and phonon properties reveals that
half-metallic K$_{2}$Cr$_{8}$O$_{16}$, upon cooling, undergoes
the correlation-assisted Peierls transition
to become a Mott-Peierls ferromagnetic insulator at low temperature.
\end{abstract}


\maketitle

\section{Introduction}
\begin{center}
\begin{figure}[h]
  \includegraphics[width=7.7 cm]{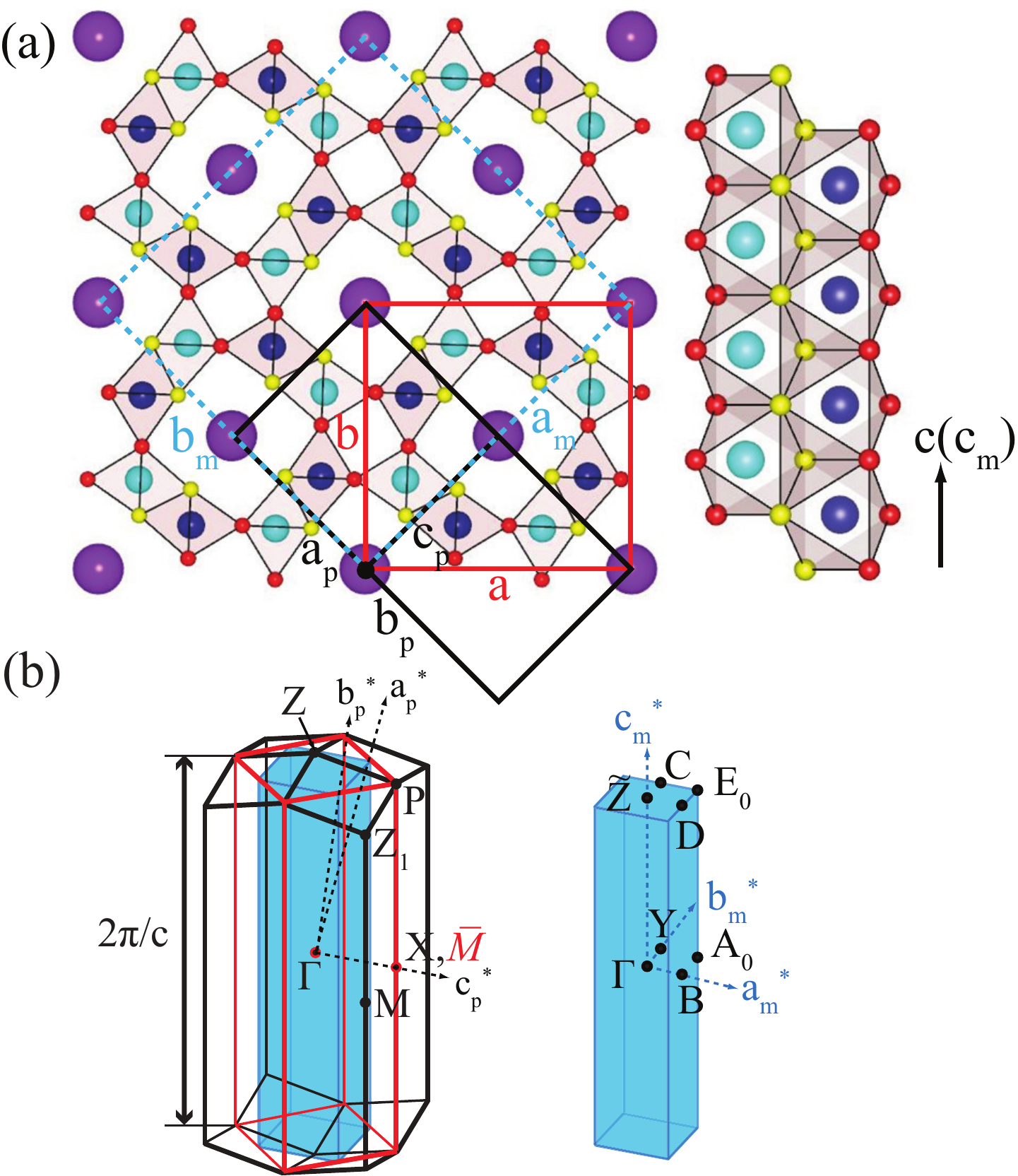}
  \caption{(Color online)
  (a) Crystal structure of K$_{2}$Cr$_{8}$O$_{16}$.
  Black and red lines represent the primitive and the conventional unit cell of the FM phase,
respectively, while dotted blue represents the unit cell of the FI phase.
  Purple, blue, sky-blue, yellow, and red balls represent K, Cr(z=0), Cr(z=0.5)
   O(1), and O(2), respectively.
  Right is the double-chain of edge-shared CrO$_{6}$ octahedra along the $c$ axis.
  In the FM phase, there are two types of O sites, edge-shared O(1) and corner-shared O(2).
  (b) The BZ of K$_{2}$Cr$_{8}$O$_{16}$.
Black and red lines represent the BZ's of the primitive and the conventional cell of the FM phase,
respectively, and the blue filled box represents the BZ of the FI phase.
($a_p^*, b_p^*, c_p^*$) and ($a_m^*, b_m^*, c_m^*$) are reciprocal lattice vectors of
the primitive body-centered tetragonal unit cell of the FM phase
and the monoclinic unit cell of the FI phase, respectively.
Note that, in the FM phase, $\bar{M}$
 of the BZ of the conventional cell is equivalent
to X of the BZ of the primitive cell.
}\label{str}
\end{figure}
\end{center}

Strongly correlated transition-metal (TM) oxides
have been studied for last decades because of their diverse intriguing properties
according to pressure, doping, temperature, and so on.\cite{Imada98}
Among those, hollandite-type TM oxides
{\it A}$_{2}${\it M}$_{8}$O$_{16}$ ({\it A}= alkali metal, {\it M}= TM element)
have drawn recent attention due to their 
quasi-one-dimensional column structure made of four double
$M$O chains (see Fig. \ref{str}).
Despite the similarity in the crystal structures, these materials have different
physical properties depending on the TM element.
K$_{2}$Cr$_{8}$O$_{16}$ and K$_{2}$V$_{8}$O$_{16}$ exhibit
the metal-insulator transition (MIT) and the structural transition concomitantly
upon cooling.\cite{Toriyama11,Isobe06}
In contrast, K$_{2}$Mo$_{8}$O$_{16}$ is an insulator
for the entire temperature range.\cite{Ozawa06}
K$_{2}$Cr$_{8}$O$_{16}$ and K$_{x}$Ti$_{8}$O$_{16}$ show magnetic transition
from paramagnetic to ferromagnetic (FM) phase
with varying the temperature and the carrier concentration,
respectively.\cite{Hasegawa09,Naomi10,Isobe09}
On the other hand, K$_{2}$V$_{8}$O$_{16}$ has a complex ground state
containing the charge ordering and the spin-singlet structure.\cite{Isobe06,Horiuchi08}
The possibility of Tomonaga-Luttinger liquid property has also been
suggested in K$_{2}$Ru$_{8}$O$_{16}$.\cite{Toriyama2011}

K$_{2}$Cr$_{8}$O$_{16}$ of present concern exhibits an interesting phase diagram
depending on the temperature (T).
It would have nominal mixed-valence state with the average electrons per Cr being 2.25
(Cr$^{4+}$ : Cr$^{3+}$ = 3:1).
At high T, K$_{2}$Cr$_{8}$O$_{16}$ is a paramagnetic metal
with the body-centered tetragonal crystal structure ({\emph I}4/{\emph m}).\cite{Hasegawa09,Tamada96}
Figure~\ref{str} presents the crystal structure and the corresponding Brillouin zone (BZ).
Along the {\it c} axis, the double-chain of edge-shared CrO$_{6}$ octahedra exists.
Upon cooling, two phase transitions take place.
At first, the magnetic transition occurs at 180K from paramagnetic to FM phase.
The observed magnetic moment of the system is close to 18 $\mu_{B}$.\cite{Endo76,Hasegawa09}
This magnetic transition is not accompanied with the structural transition,
and so this phase is still metallic.
Interestingly,
the FM phase has half-metallic nature.\cite{Hasegawa09,Sakamaki09}
This feature was explained by the double exchange mechanism
arising from the Hund coupling between the fully occupied {\it d$_{xy}$} orbital
and the partially occupied {\it d$_{zx}$}
and {\it $d_{yz}$} of Cr,\cite{Sakamaki09}
as in CrO$_{2}$.\cite{Korotin98}

The second transition is the MIT at 95 K.
At this transition, the ferromagnetic property is preserved.\cite{Hasegawa09}
An insulator version of the double exchange ferromagnetism was proposed
to explain this property.\cite{Nishimoto12}
When the MIT was first observed, the structural transition was not detected,\cite{Hasegawa09}
and so the charge ordering\cite{Mahadevan10} and the charge density wave
\cite{Sakamaki09} were suggested as the mechanism of the MIT.
Later, however, the structural transition coexisting with the MIT was
observed,\cite{Toriyama11} from the body-centered tetragonal to the monoclinic structure
({\emph P}112$_{1}$/{\emph a}) that has an increased unit cell of
$\sqrt{2} \times \sqrt{2} \times 1$.
Based on this newly discovered structure, it was suggested that
the MIT is caused by a Peierls instability in the quasi-one-dimensional column structure
made of four coupled Cr-O chains running along the $c$ direction.\cite{Toriyama11}

There have been a few electronic band structure studies on
K$_{2}$Cr$_{8}$O$_{16}$.\cite{Sakamaki09,Mahadevan10,Toriyama11}
By contrast, there has been no phonon study, which can provide a direct clue to
the structural transition.
Since the electronic structure of K$_{2}$Cr$_{8}$O$_{16}$ is similar to that of CrO$_2$,
it is expected that the Coulomb correlation effect plays an important role in
determining physical properties.
However, the roles of the Coulomb correlation in the MIT and in stabilizing the
ferromagnetic insulator (FI) ground state have not been explored yet.

In this article, we have investigated electronic structures and the phonon dispersion
properties of K$_{2}$Cr$_{8}$O$_{16}$
to explore the driving mechanism of the MIT and the structural transition.
To analyze the effect of on-site Coulomb correlation, $U$,
we studied the phonon properties both without and with $U$ in the density functional theory (DFT)
calculations.
We have found that the structural instability
occurs only in the DFT+$U$, which indicates that the Coulomb correlation plays an
essential role in the structural transition.
The displacements of normal mode of softened phonon at X (0, 0, 1/2) are quite consistent with
the observed structural distortions occurring at the MIT,
which leads to the formation of Cr tetramers
in the four coupled Cr-O chains and their stripe-type arrangement.
Noteworthy is that this {\bf Q} = X (0, 0, 1/2) is different from the previously
suggested Peierls nesting vector of {\bf Q}= (1/2, 1/2, $-1/2$).\cite{Toriyama11}
We have demonstrated that the FI phase of K$_{2}$Cr$_{8}$O$_{16}$ belongs to
a Mott-Peierls insulator, for which both
the lattice displacements and the Coulomb correlation are essential in opening the gap.

\section{Computational Details}

For the DFT calculations,
we employed the pseudo-potential band method implemented in VASP,\cite{Kresse96}
and for the phonon calculations, we employed PHONOPY code.\cite{Togo08}.
The force constants for the phonon calculation were obtained from the supercell
calculation with finite displacements by the Hellmann-Feynman theorem.\cite{Parlinski97}
We used the pseudo-potential band method with projected augmented-wave (PAW) potentials.\cite{Blochl94,Kresse99}
The generalized gradient approximation (GGA-PBE)
was used for the exchange-correlation functional.\cite{Perdew96}
Before carrying out the phonon calculations, we performed the full-relaxation, which
adjusts the lattice constant and atomic positions
under the symmetry constraint of the original structure.
The initial structural data before the full-relaxation
were taken from the experiment.\cite{Nakao12}

For the GGA+$U$ calculations, we have checked several $U$ values,
and presented the results for $U= 3.0$ eV below.\cite{Toriyama11}
The energy cut off of 520 eV was used for the number of plane wave bases.
The {\bf k}-point samplings are (6$\times$6$\times$20) for the conventional
tetragonal unit cell of the FM metallic phase,
and (4$\times$4$\times$20) for the monoclinic unit cell
of the FI phase in the Monkhorst-Pack grid.
The (2$\times$2$\times$2) supercell of
the conventional tetragonal structure
is selected for the phonon calculation of the FM metallic phase.

\section{Results and Discussions}

We first carried out the band structure calculations for the FM phase
of K$_{2}$Cr$_{8}$O$_{16}$ both in the GGA and the GGA+$U$ schemes.
Both schemes produces the metallic phase,
more precisely, the half-metallic phase,
as is consistent with existing reports.\cite{Sakamaki09,Hasegawa09}
In both cases, total magnetic moments per formula unit are obtained to be 18.00 $\mu_{B}$,\cite{localM}
which is close to the experimental value 
{\bf \cite{Endo76,Hasegawa09}}

\begin{center}
\begin{figure}[t]
  \includegraphics[width=8 cm]{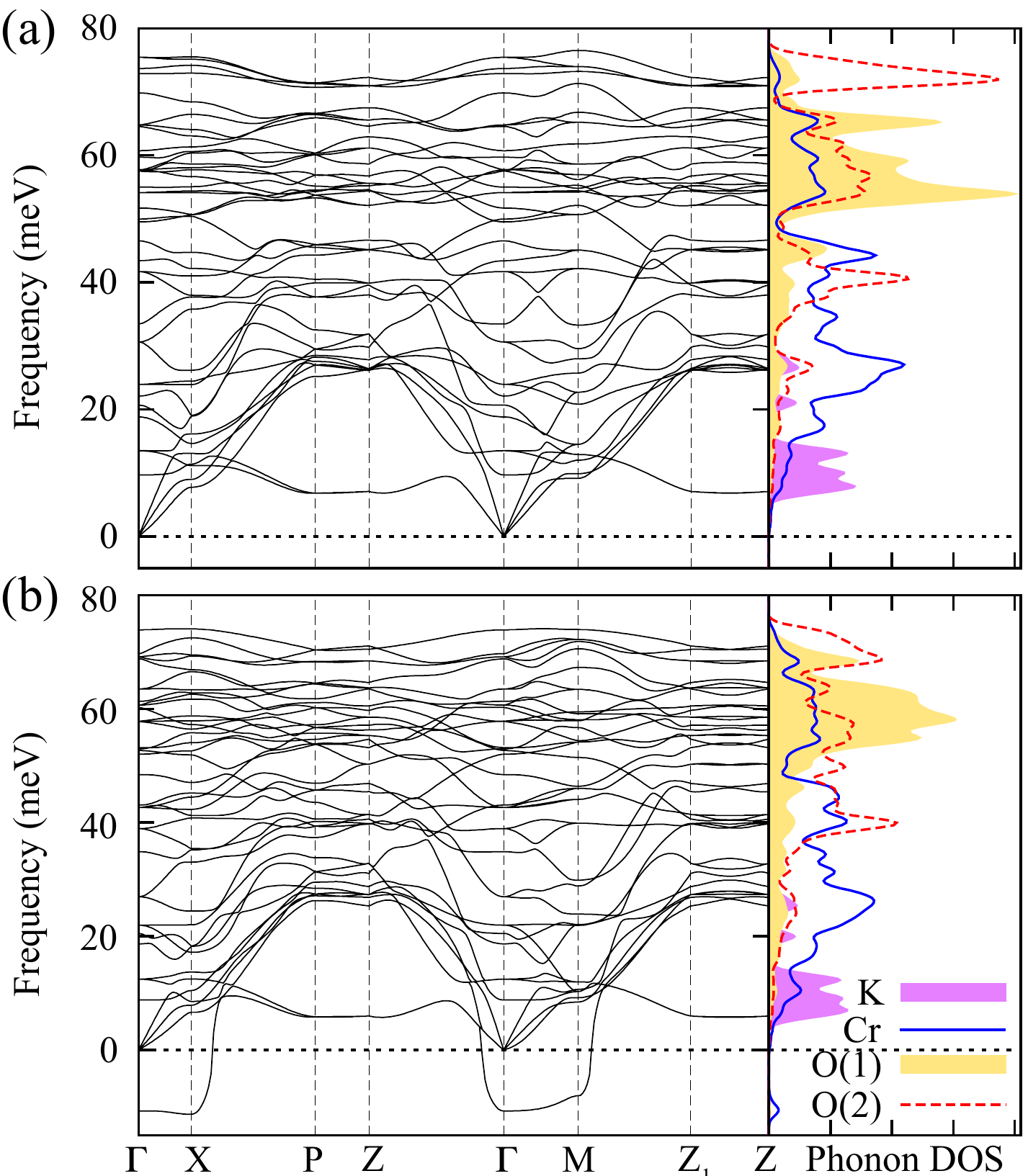}
  \caption{(Color online)
 Phonon dispersion and the corresponding partial DOS of K$_{2}$Cr$_{8}$O$_{16}$
	in its FM phase :
(a) result from the GGA, and (b) result from the GGA+$U$ ($U= 3.0$ eV).
  The k-point symmetry lines are along the BZ boundaries of primitive unit cell of the FM phase.
  The negative phonon frequencies (imaginary frequencies) reflects
  the structural instability. 
It is seen from the DOS in (b) that the softened phonons are related
mostly to the lattice vibrations of Cr ions.
}
\label{phd}
\end{figure}
\end{center}

Figure \ref{phd} presents the phonon dispersion curve and the phonon partial density of
states (DOS) of K$_{2}$Cr$_{8}$O$_{16}$ in its FM phase.
Figure~\ref{phd}(a) and (b) were calculated without and with $U$, respectively.
In both cases, flat bands are observed at $\sim 10$ meV, which correspond to the
localized phonon bands from K.
The significant difference between phonon bands of the GGA and the GGA+$U$ is
the existence of the imaginary phonon frequency in the latter (Fig. \ref{phd}(b)),
which indicates the structural instability of the FM phase
of K$_{2}$Cr$_{8}$O$_{16}$ in agreement with the experiment.
The fact that the structural instability is obtained in the GGA+$U$
but not in the GGA reflects that
the GGA+$U$ phonon calculation describes the experimental structural transition properly,
while the GGA phonon calculation does not.
It is seen that the structural instabilities are prominent at M, X, and $\Gamma$,
which correspond to mostly Cr displacements as shown in phonon DOS.
We also checked the structural instability with varying $U$ value.
It occurs only when $U > \sim$2.5eV.
This means that large enough $U$ is necessary to induce the structural instability,
which corroborates that the structural transition is really driven by the Coulomb correlation effect.
We will discuss below that the softened phonon responsible for the transition
in K$_{2}$Cr$_{8}$O$_{16}$ is that at X.

To investigate the structural instability signaled by phonon in more detail, 
let us examine the low temperature structure of the FI phase of K$_{2}$Cr$_{8}$O$_{16}$.
The main difference between the FM phase and the FI phase is the increased unit cell size
by $\sqrt{2} \times \sqrt{2} \times 1$ with respect to
the conventional cell of the FM phase due to the distortions of Cr and O.
Toriyama {\it et al}.\cite{Toriyama11} observed that, in the {\it ab} plane,
there is a stripe-type arrangement of the Cr tetramers formed in the four-chain columns.
To explain the modulation of the tetragonal cell in the {\it a} and {\it b} directions,
there should be a structural instability at $\bar{M}$ in the conventional tetragonal cell,
which actually corresponds to the X point in the primitive cell of the FM phase
(see Fig. \ref{str}(b)).
It is because
the softened phonon mode at $\bar{M}$ in the tetragonal unit cell
exhibits the opposite atomic displacements alternating in every other unit cell
along the {\it a} and {\it b} directions and increases the unit cell in {\it ab} plane.
Note that this {\bf Q}(=X) vector is different from the previously
suggested Peierls nesting vector Q$_{z}$=2$\pi/c$,\cite{Toriyama11}
which actually corresponds to M point (1/2, 1/2, $-1/2$)
in the primitive cell or $\Gamma$ point in the conventional cell.

\begin{center}
\begin{figure}[t]
  \includegraphics[width=9.0 cm]{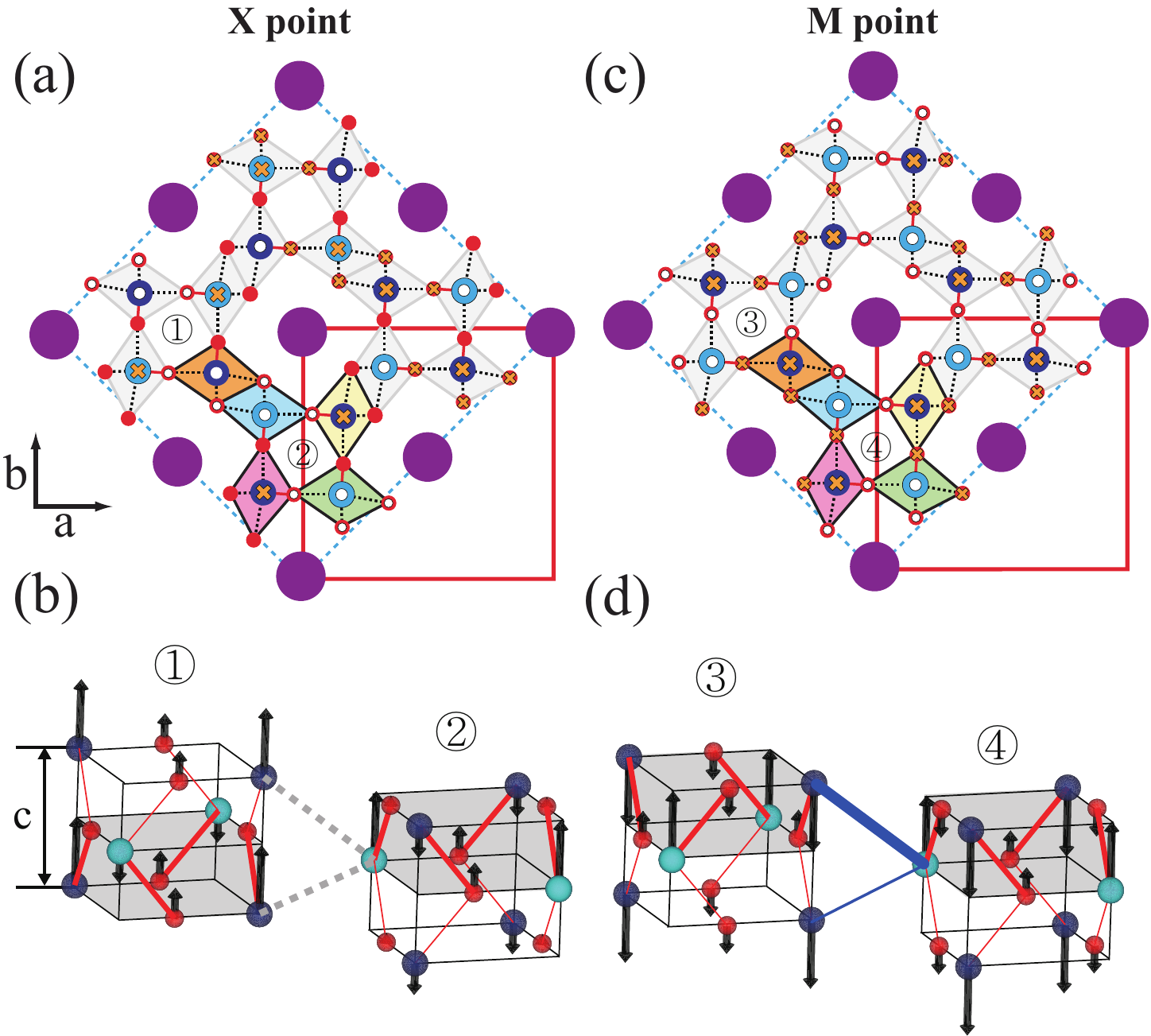}
  \caption{(Color online)
  (a),(b) Lattice displacements of the softened phonon mode at X,
and (c),(d) those at M.
 Purple, blue, sky-blue, and red circles are K, Cr(z=0), Cr(z=0.5) and O respectively.
  White circles and orange crosses at Cr and O
   in (a) and (c) denote the outward and inward arrows, respectively.
 (b) and (d) show the main atomic displacements of Cr and O(2),
	which lead to the formation of Cr tetramers in the four-chain column,
	as depicted by gray boxes in \textcircled{1}, \textcircled{2}, \textcircled{3}, and \textcircled{4}.
Thick and thin red lines represent shorter and longer bonds between Cr and O ions,
respectively.
Dotted gray lines connecting Cr ions in-between the double-chain of (b)
stand for the uniform distance between Cr-Cr, whereas thick and thin blue lines in (d)
for the shorter and longer distances between Cr-Cr, respectively.
}
\label{dis}
\end{figure}
\end{center}
\begin{center}
\begin{figure}[b]
  \includegraphics[width=8.0 cm]{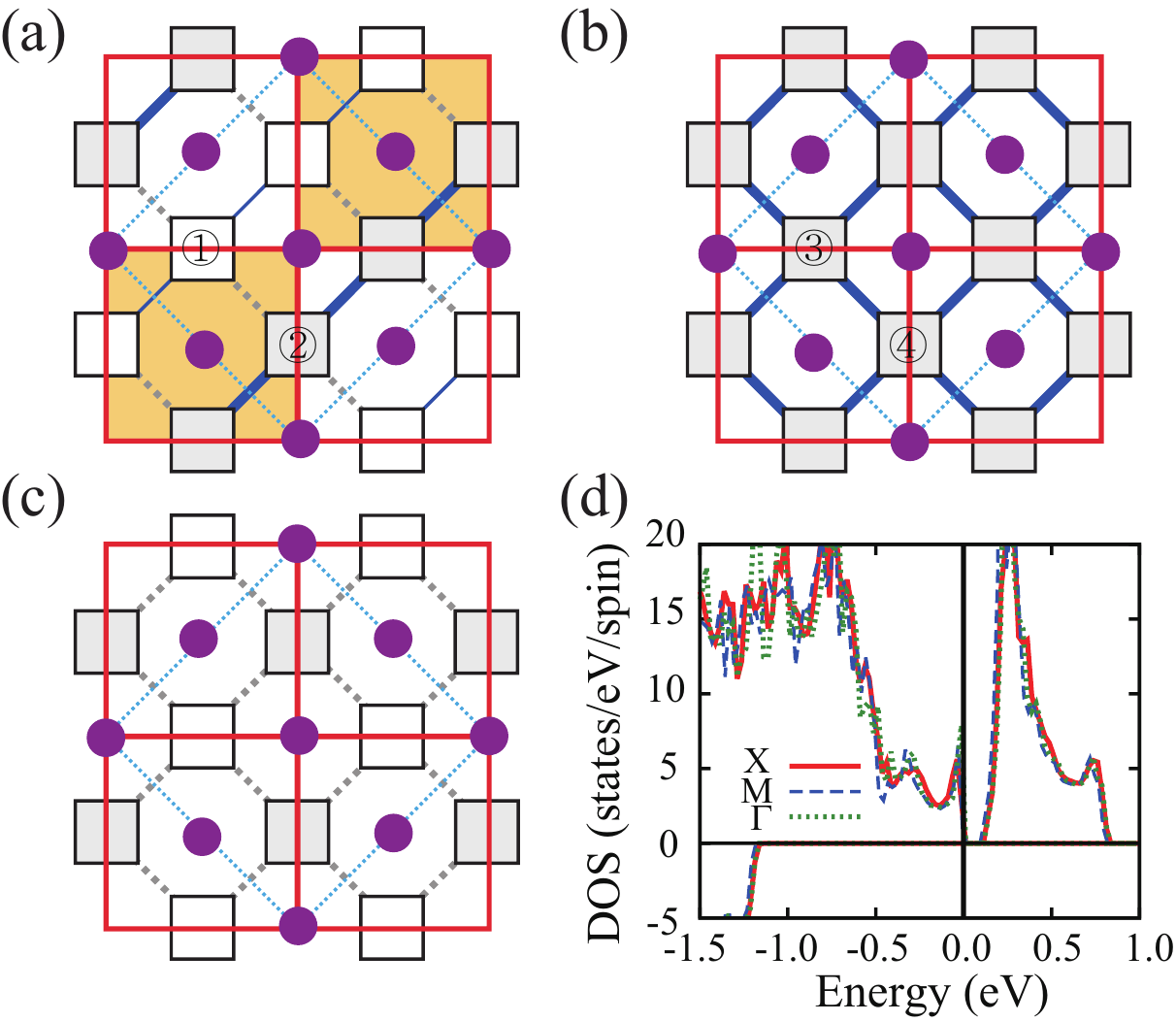}
  \caption{(Color online)
  The arrangements of Cr tetramers in the {\it ab} plane obtained from the softened phonon modes.
  Dotted gray, thick and thin blue lines connecting Cr ions in the double-chains are
	the same as in Fig. \ref{dis}.
Solid purple circles represent K atoms.
 (a) The stripe-type arrangement by the softened phonon at X. Yellow shadow indicates
 the modulation of the tetragonal structure in the {\it ab} plane,
as observed in the experiment.\cite{Toriyama11}
 (b) The uniform-type arrangement by the softened phonon at M.
 (c) The checkerboard-type arrangement by the softened phonon at $\Gamma$.
 (d) Total DOS of each tetramer arrangement obtained in the GGA+$U$ ($U= 3.0$ eV).
	Positive and negative DOSs represent those of spin-majority and spin-minority
	electrons, respectively.
}
\label{XMG}
\end{figure}
\end{center}

Figure \ref{dis} shows the lattice displacements of softened phonon modes at X and M.
In both cases, the formation of Cr tetramers, which is the main lattice distortion
from FM to FI phase, appears, as depicted by gray boxes in
\textcircled{1}, \textcircled{2}, \textcircled{3}, and \textcircled{4} in Fig. \ref{dis}(b) and (d).
The softened phonon at X in Fig. \ref{dis}(a) and (b) produces not only the formation
of Cr tetramer in the four-chain column along the {\it c} direction but also
the stripe-type arrangement of Cr tetramers in the {\it ab} plane, as shown in
Fig. ~\ref{XMG}(a).
The formation of Cr tetramers shown in Fig. ~\ref{dis}(b) is
perfectly consistent with that reported by Toriyama {\it et al.}.\cite{Toriyama11}
The stripe-type arrangement of Cr tetramers explains the unit cell increasing
in the {\it ab} plane after the MIT.
Namely, the normal modes of softened phonon at X is exactly consistent
with the experimental lattice distortions.

Figure~\ref{dis}(c) and (d) present the lattice displacements by the softened phonon at M,
which is the previously suggested Peierls nesting vector,
Q$_{z}$=2$\pi/c$.\cite{Toriyama11}
The formation of Cr tetramers is also obtained in this case, as shown in Fig. ~\ref{dis}(d).
However, the arrangement of Cr tetramers in the {\it ab} plane is different
from the experimentally observed stripe-type.
In this case, the arrangement is of uniform-type, as shown in Fig. \ref{XMG}(b),
which can not explain the increasing unit cell from the FM phase to the FI phase.
This results is somehow expected because
M point in the primitive cell of the FM phase is the same position as $\Gamma$ point in the
conventional tetragonal cell of the FM phase.
That is why the structural instability at M produces the uniform-type arrangement of Cr tetramers,
which does not increase the unit cell of the conventional tetragonal structure
in the {\it ab} plane.
We thus suggest that the Peierls distortion
vector {\bf Q} that yields the Peierls instability is
X (0, 0, 1/2) in the primitive cell of the body-centered tetragonal structure,
not M (1/2, 1/2, $-1/2$).

This result is indeed consistent with that from total energy calculations.
We checked the total energies of modulated structures corresponding
to the softened phonon modes at X, M, and $\Gamma$ with varying associated displacements.
At each mode, the total energy has the minimum at the equilibrium displacement.
It is found that the total energy of the modulated structure due to X phonon mode is the lowest,
while that due to M phonon mode is the highest. 
This result is in agreement with the above phonon mode analysis.

The main difference between the lattice displacements of softened phonons at X and M
lies in the linking of Cr-Cr in the double-chains of the edge-shared CrO$_{6}$ octahedra.
If Cr atoms in each chain move oppositely,
the Cr-Cr dimerization of zig-zag type is formed in the double-chain,
as shown in Fig. \ref{dis}(d).
Otherwise, there is no Cr-Cr dimerization (Fig. \ref{dis}(b)).
Due to these two possibilities of Cr-Cr linkings in the double-chains,
there occur three types of arrangements of Cr tetramers,\cite{Nakao12} as shown in Fig. ~\ref{XMG}.
The stripe-type in Fig. \ref{XMG}(a)
contains the half of double-chains with Cr dimers,
while the uniform-type in Fig. \ref{XMG}(b) and the checkerboard-type in Fig. \ref{XMG}(c)
have all and none of the double-chains with Cr dimers, respectively.
Noteworthy is that these three types of arrangements match exactly with
the normal mode displacements of softened phonons at X, M and $\Gamma$.

To check the role of the Cr-Cr dimers in the double-chains,
we compared the DOSs of the above three modulated structures in Fig. \ref{XMG}(d).
Intriguingly, the insulating phases are obtained for all three cases,
regardless of existence or non-existence of
the Cr-Cr dimers in the double-chain.
The sizes of energy gaps of three cases are 0.1$\sim$0.13 eV, which are
similar to that of the FI phase.\cite{Toriyama11}
This result indicates that the insulating nature of the FI phase originates from
the formation of Cr tetramers in the four-chain columns,
not from the Cr-Cr dimers in the double-chains.
This feature is quite different from the case in K$_2$Mo$_8$O$_{16}$, for which
the insulating gap was reported to arise from the distortions in the double-chain
of the edge-shared MoO$_6$ octahedra.\cite{Toriyama13}

\begin{center}
\begin{figure}[t]
  \includegraphics[width= 8.5 cm]{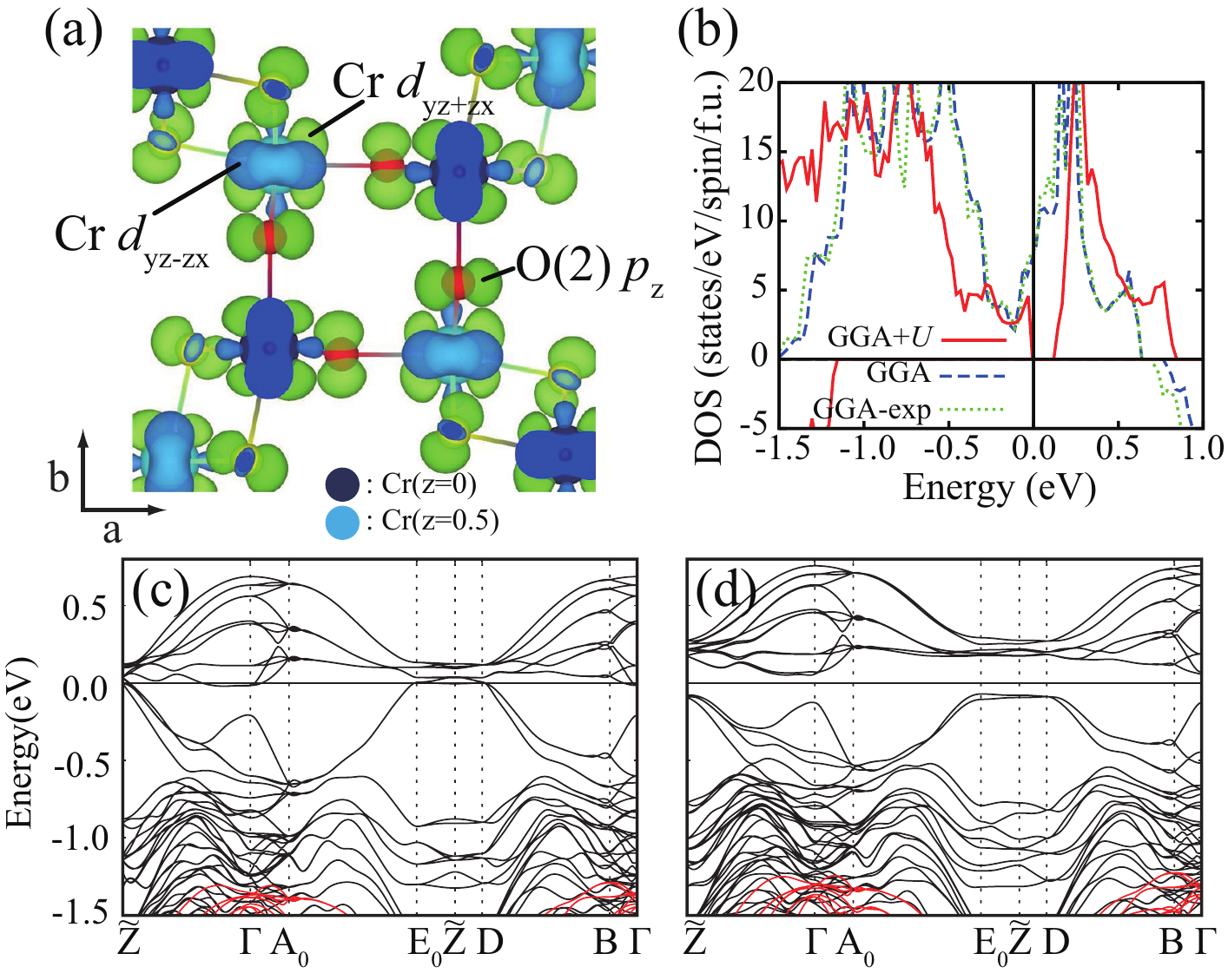}
  \caption{(Color online)
   (a) The CDD between the GGA+$U$ ($U= 3.0$ eV) and the GGA for the FM phase of K$_{2}$Cr$_{8}$O$_{16}$.
   Green and blue represent the positive and negative part, respectively.
   The Cr tetramer, indicated by the red line, is seen to be connected by the $\pi$-bonding.
   (b) Comparison of total DOSs of the GGA and the GGA+$U$ ($U= 3.0$ eV) for the FI phase of K$_{2}$Cr$_{8}$O$_{16}$.
The DOS of GGA-exp is obtained for the experimental structure in the GGA.
 (c) The GGA+$U$ ($U= 3.0$ eV) band structure of the FM phase of K$_{2}$Cr$_{8}$O$_{16}$, and
 (d) that of the FI phase of K$_{2}$Cr$_{8}$O$_{16}$,
	both of which are plotted in the BZ of the FI phase.
Band splitting is evident in the BZ boundary including $\tilde{Z}$.
  Spin-majority and minority bands are plotted by black and red lines, respectively.
The structures of FM and FI phases were optimized before the calculations
except the case of GGA-exp in (b).
}
\label{dos}
\end{figure}
\end{center}

We also checked the roles of the Coulomb correlation $U$ and the Peierls dimerization
in the MIT and the structural transition.
To investigate the $U$ effect in the FM phase, the charge density difference (CDD)
between the GGA+$U$ and GGA calculations is plotted in Fig. ~\ref{dos}(a).
It is seen that Cr {\it d$_{yz+zx}$} and O(2) {\it p$_{z}$} electrons are more occupied
in the GGA+$U$ than in the GGA,  while Cr {\it d$_{yz-zx}$} electrons are less occupied.
This result suggests that the hybridization $\pi$-bonding between
Cr {\it d$_{yz+zx}$} and O {\it p$_{z}$} orbitals increases with $U$.
In fact, the Cr {\it d$_{yz+zx}$}-O {\it p$_{z}$} spin transfer path
in the FI phase is emphasized in the recent NMR paper.\cite{Takeda13}
This hybridization is important for the formation of Cr tetramer,
because Cr ions in tetramer are linked with O(2).
Therefore, $U$ effect facilitates the transition from the FM phase to the FI phase
by increasing the hybridization between Cr {\it d$_{yz+zx}$} and O(2) {\it p$_{z}$}.
Thus, this increased hybridization is expected to cause the structural instability
in the phonon dispersion of the FM phase, so as to produce the stabilized Cr tetramers.
Phonon data and the CDD result are reminiscent of the correlation assisted Peierls transition
or Mott-Peierls transition in VO$_{2}$.\cite{Biermann05,Haverkort05,Kim13}
But the unique feature of K$_{2}$Cr$_{8}$O$_{16}$ is that the Mott-Peierls transition
occurs in the fully spin-polarized band.

Figure \ref{dos}(b) shows the role of $U$ in the FI phase in the band gap opening.
The band gap opens only when both the distortion and the Coulomb correlation are
taken into account.
We found that the Cr-tetramer distortion is stabilized only with the Coulomb correlation.
In the GGA calculation, a Cr tetramer given for an initial structure disappears
after the structural optimization, and there is no band-gap opening.
The GGA calculation for the experimental structure of the FI phase containing Cr tetramers
also does not produce the band gap
(see the DOS of GGA-exp in Fig. \ref{dos}(b)).
Namely, the GGA calculation does not open the band gap,
even in the presence of the Peierls dimerization.
This implies that K$_{2}$Cr$_{8}$O$_{16}$ belongs to a Mott-insulator
possessing the Peierls-type instability.
The band gap in the GGA+$U$ is 0.125 eV, which is in agreement
with the previous calculation.\cite{Toriyama11}

Figure~\ref{dos}(c) and (d) show GGA+$U$ band structures of the FM and FI phases
of K$_{2}$Cr$_{8}$O$_{16}$, calculated in the FI phase symmetry.
In the band structure of the FM phase in Fig. \ref{dos}(c), there is no band gap opening
because of the absence of the Peierls distortion.
That is, the band gap opens
 in the presence of the Peierls distortion (Fig. \ref{dos}(d)).
Through the lattice distortion, the energy gain is obtained mainly from bands at $\tilde{Z}$,
E$_{0}$, and D, which are in the BZ boundary along the {\it c$^{*}$} direction
(see Fig. \ref{str}(b)).
The energy gain from the lattice distortion in this case is $\sim$ 19 meV/f.u..

\section{Conclusion}

In conclusion, based on the systematic studies of electronic and phonon properties,
we have found that K$_{2}$Cr$_{8}$O$_{16}$ is a strongly correlated system
having the Coulomb correlation effect both in the FM and the FI phases.
We demonstrated that the structural instability
that occurs at X only in the GGA+$U$ scheme
describes perfectly well the experimentally observed lattice distortions,
such as
the formation of Cr tetramers along the {\it c} direction and the stripe-type
arrangement of Cr tetramers in the {\it ab} plane.
Thus we proposed the Mott-Peierls transition as the driving mechanism
of the concomitant MIT and the structure transition in K$_{2}$Cr$_{8}$O$_{16}$.
Our finding of the Peierls distortion
vector {\bf Q}=X rules out the previously suggested
nesting vector of Q$_{z}$=2$\pi$/{\it c}
for the Peierls transition in K$_{2}$Cr$_{8}$O$_{16}$.
We also showed that the Cr tetramers and the insulating gap of the FI phase
are stable only with Coulomb correlation,
which corroborates that the FI phase of K$_{2}$Cr$_{8}$O$_{16}$ is a Mott-Peierls insulator.

Acknowledgments$-$
This work was supported by the NRF (No. 2009-0079947, No. 2011-0025237),
the POSTECH BSRI grant, and the KISTI supercomputing center (No. KSC-2012-C2-094).
S.K. acknowledges the support from the NRF project
of Global Ph.D. Fellowship (No. 2011-0002351).


\begin{thebibliography}{99}
\bibitem{localM}{Local magnetic moments in the GGA are 2.32$\mu_{B}$/Cr, $-0.02\mu_{B}$/O(1),
and $-0.11\mu_{B}$/O(2), while those in the GGA+$U$ are 2.59$\mu_{B}$/Cr, $-0.10\mu_{B}$/O(1), and
$-0.23\mu_{B}$/O(2).}
\end{thebibliography}
\end{document}